\documentclass{article}
\usepackage{arxiv}

\usepackage[utf8]{inputenc} 
\usepackage[T1]{fontenc}    
\usepackage{hyperref}       
\usepackage{url}            
\usepackage{booktabs}       
\usepackage{amsfonts}       
\usepackage{nicefrac}       
\usepackage{microtype}      
\usepackage{lipsum}
\usepackage{graphicx}
\graphicspath{ {./images/} }
\usepackage{booktabs}  

\usepackage[round]{natbib} 
\usepackage{xcolor}
\usepackage{caption}
\usepackage{hyperref}
\usepackage{cleveref}
\crefname{figure}{Fig.}{Figs.}
\hypersetup{
    colorlinks=true,
    linkcolor=blue,    
    citecolor=blue,    
    urlcolor=blue      
}

\newcommand{\RN}[1]{\uppercase\expandafter{\romannumeral#1}}

\title{Automating Structural Analysis Across Multiple Software Platforms Using Large Language Models}

\author{
Ziheng Geng$^{1}$,
Jiachen Liu$^{2}$,
Ian Franklin$^{1}$,
Ran Cao$^{3}$,
Dan M. Frangopol$^{4}$,
Minghui Cheng$^{1,5\dagger}$\\
\\
$^{1}$Department of Civil and Architectural Engineering, University of Miami, Coral Gables, FL 33146, USA\\
$^{2}$HBC Engineering Company, Miami, FL 33178, USA\\
$^{3}$College of Civil Engineering, Hunan University, Changsha, 410082, China\\
$^{4}$Department of Civil and Environmental Engineering, Lehigh University, Bethlehem, PA 18015, USA\\
$^{5}$School of Architecture, University of Miami, Coral Gables, FL 33146, USA\\
\\
$^{\dagger}$Corresponding author: \texttt{minghui.cheng@miami.edu}
}

\begin{document}
\maketitle
\begin{abstract}
Recent advances in large language models (LLMs) have shown the promise to significantly accelerate the workflow by automating structural modeling and analysis. However, existing studies primarily focus on enabling LLMs to operate a single structural analysis software platform. In practice, structural engineers often rely on multiple finite element analysis (FEA) tools, such as ETABS, SAP2000, and OpenSees, depending on project needs, user preferences, and company constraints. This limitation restricts the practical deployment of LLM-assisted engineering workflows. To address this gap, this study develops LLMs capable of automating frame structural analysis across multiple software platforms. The LLMs adopt a two-stage multi-agent architecture. In Stage 1, a cohort of agents collaboratively interpret user input and perform structured reasoning to infer geometric, material, boundary, and load information required for finite element modeling. The outputs of these agents are compiled into a unified JSON representation. In Stage 2, code translation agents operate in parallel to convert the JSON file into executable scripts across multiple structural analysis platforms. Each agent is prompted with the syntax rules and modeling workflows of its target software. The LLMs are evaluated using 20 representative frame problems across three widely used platforms: ETABS, SAP2000, and OpenSees. Results from ten repeated trials demonstrate consistently reliable performance, achieving accuracy exceeding 90\% across all cases.

\end{abstract}

\begin{quote}
\end{quote}

\section{Introduction}
Powered by billions of parameters, state-of-the-art large language models (LLMs), such as GPT-5 \citep{openai2025gpt52update}, Gemini 3 \citep{google2025gemini3procard}, and DeepSeek V3 \citep{liu2024deepseek}, have demonstrated strong generalization capabilities across tasks including code generation \citep{jiang2024survey, joel2024survey, dong2025survey}, long-context reasoning \citep{bai2025longbench, chung2025evaluating, ling2025longreason}, and tool calling \citep{liu2024toolace, shen2024llm, lu2025toolsandbox}. This technological leap offers a unique opportunity to leverage these capabilities to automate intricate workflows and accelerate the efficiency of civil engineers. Initial explorations into LLM applications have successfully employed these models for tasks such as building code interpretation \citep{joffe2025framework,zhu2024llm,chen2025chatcivic}, construction report generation \citep{pu2024autorepo, chen2025meet2mitigate, xia2026graph}, and visual damage assessment \citep{zhang2025sdiglm,jiang2025large,jiang2026multitask}. Collectively, these studies underscore the transformative potential of LLMs to augment human expertise and assist in complex engineering tasks.

Within civil engineering, structural modeling and analysis is one of the cornerstones for the design and assessment of civil infrastructure systems. In contemporary practice, the finite element analysis (FEA) has become indispensable to analyze structural behavior under complex geometric, loading, and material conditions. A broad range of FEA software platforms have been widely adopted by engineers, including commercial software such as ETABS \citep{etabs2023} and SAP2000 \citep{sap2000}, as well as open-source frameworks such as OpenSees \citep{mckenna2011opensees}. In practice, using FEA software is largely manual and time-consuming. While LLMs possess the potential to automate these workflows, they lack the native ability to interact iteratively with external computational environments. To bridge this gap, recent research has shifted toward the development of LLMs capable of using domain software.  \cite{liu2026large} and \cite{geng2025lightweight, geng2026novel} adapted lightweight Llama 3.3 70B Instruct models to generate OpenSeesPy code to automate structural analysis of beam and frame structures. \cite{liang2025integrating, liang2025automating} proposed domain-specific prompt design for structural engineering problems and demonstrated their approach through frame analysis and racking system design. Focusing on the design, \cite{chen2026multi} developed LLMs that automate ABAQUS operations to identify optimal design solutions of ultra-high-performance concrete beams. In addition, \cite{deng2025bimgent} and \cite{du2026text2bim} trained LLMs to use building information modeling softwre to construct simple 3D building models from natural language instructions and conceptual sketches. 

\begin{figure*}[htbp]
\centering
\includegraphics[width=0.85\textwidth]{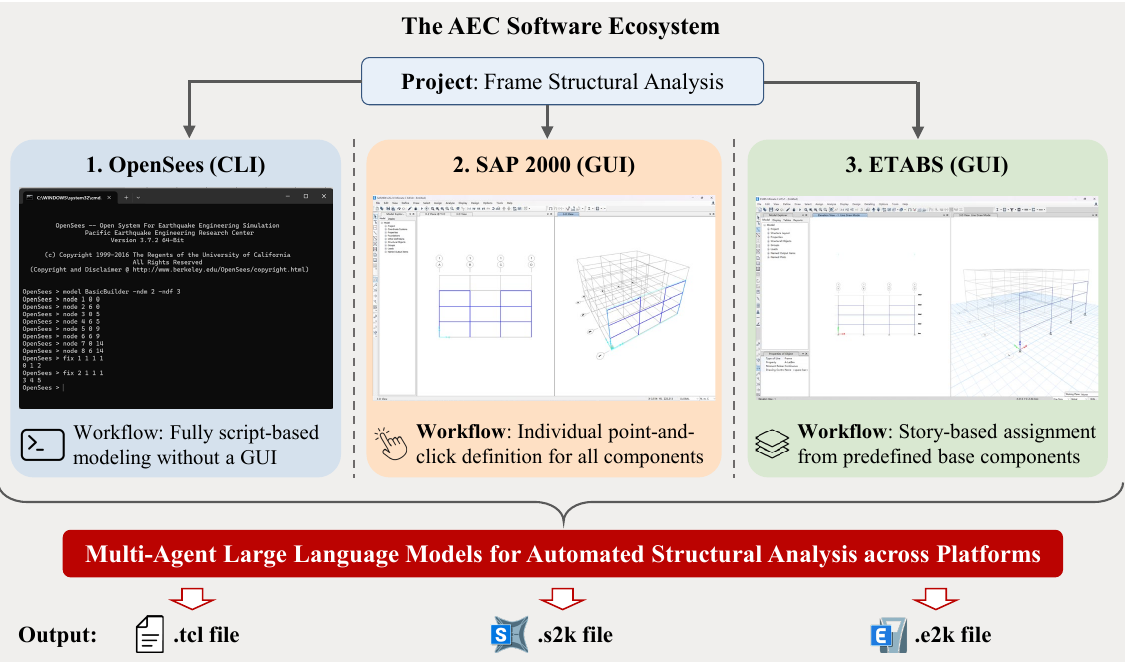}
\caption{Development of LLMs capable of operating multiple software platforms for automated structural analysis.}
\label{Figure1}
\end{figure*}

While existing studies are pushing the boundary in training LLMs to use engineering software, there is a lack of research in developing capabilities to operate different software platforms. In practice, engineering workflows rarely rely on a single tool, and multiple FEA platforms (such as ETABS, SAP2000, and OpenSees) are commonly used within the structural engineering community depending on the project needs and constraints. Therefore, developing LLMs capable of operating multiple platforms is essential for practical LLM-assisted engineering. Otherwise, separate LLMs would need to be trained for each software environment, significantly limiting scalability and adoption. To bridge this gap, this study develops LLMs to automate structural analysis across software silos using a multi-agent architecture, as illustrated in \cref{Figure1}. The general concept behind the architecture is that the overall task of structural modeling and analysis is decomposed into two subtasks: (i) reasoning from user input to derive the information required for finite element modeling, and (ii) translating the derived information into executable scripts tailored to specific software platforms. Following this concept, the multi-agent architecture has two stages. In Stage 1, a group of reasoning agents interprets user input and derives geometry, materials, boundary conditions, and loads, which are compiled into a unified JSON representation. In Stage 2, code translation agents convert this representation into executable scripts for different platforms. GPT-OSS 120B is used for reasoning tasks, while Llama-3.3 70B Instruct Turbo performs code translation. The system is evaluated on 20 representative frame problems across OpenSees, SAP2000, and ETABS, demonstrating reliable script generation and significantly outperforming state-of-the-art general-purpose LLMs.

\begin{figure*}[htbp]
\centering
\includegraphics[width=0.85\textwidth]{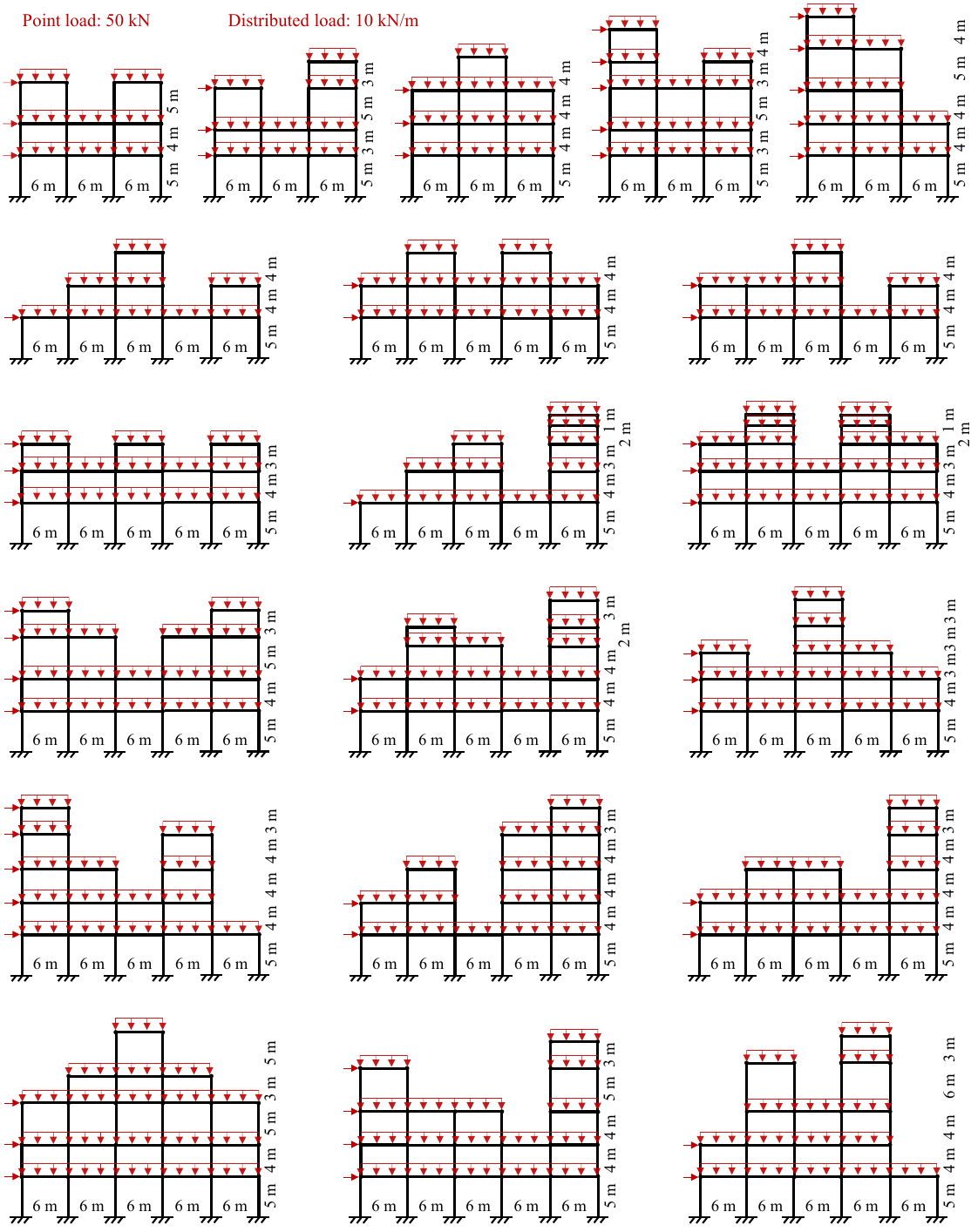}
\caption{A benchmark dataset comprising twenty representative frame structural analysis problems (adapted from \citeauthor{geng2025lightweight}, \citeyear{geng2025lightweight}).}
\label{Figure2}
\end{figure*}

\section{Benchmark Dataset}
To demonstrate how automating structural analysis across multiple software platforms using LLMs can be realized in practice, this study selects frame structural analysis as a representative case study for implementation and validation. Accordingly, a benchmark dataset comprising 20 representative frame analysis problems is adapted from previous work \citep{geng2025lightweight} and used as the evaluation testbed, as illustrated in \cref{Figure2}. The dataset encompasses a broad spectrum of geometric configurations with irregular topologies. It is generated by varying bay counts, story numbers, bay widths, and story heights to reflect realistic design scenarios. Consistent loading patterns are applied across all benchmark cases: each frame is subjected to a combination of concentrated and distributed loads, including lateral point loads applied at each floor level from the left side and a uniformly distributed gravity load applied to all beam elements. Boundary conditions and material properties are also prescribed consistently. Specifically, all base nodes are restrained to simulate fixed supports, and structural members are assumed to be linear elastic. Distinct cross-sectional properties are assigned to beams and columns, respectively. Together, these test cases serve as a systematic testbed for assessing the accuracy, robustness, and generality of the proposed LLMs.

\begin{figure*}[htbp]
\centering
\includegraphics[width=0.85\textwidth]{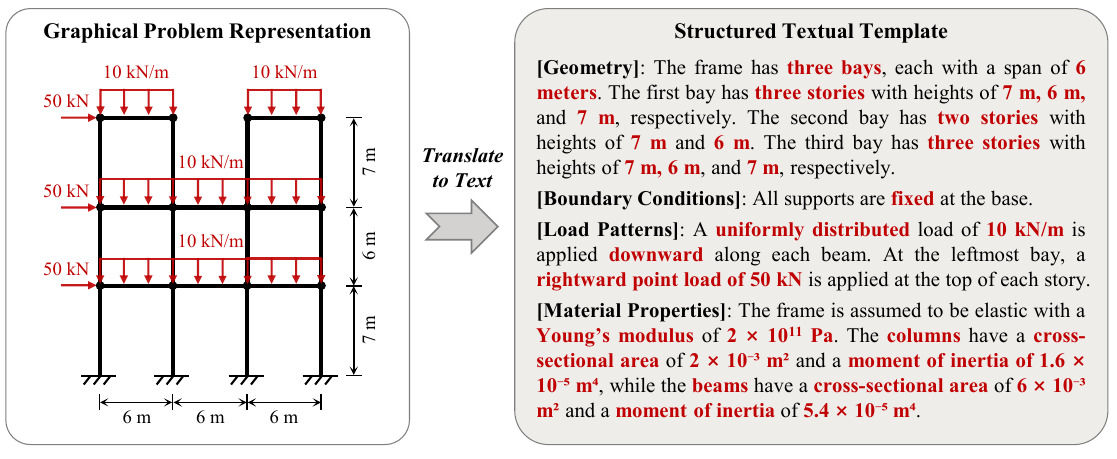}
\caption{Textual template for problem description in frame structural analysis.}
\label{Figure3}
\end{figure*}

For each benchmark problem, a structured textual description is constructed to encode the engineering intent underlying the graphical representation, following the approach proposed by \cite{wan2025som}. The template organizes problem definition into four modules: geometric configuration, boundary conditions, loading patterns, and material properties, as illustrated in \cref{Figure3}. The geometric configuration module specifies the structural topology, including the number of bays and stories, along with corresponding bay widths and story heights. Boundary conditions are defined by specifying support types and their nodal locations. Loading patterns are described in terms of load types, magnitudes, spatial directions, and application points. Material properties are prescribed through constitutive parameters required for stiffness matrix formulation, such as Young’s modulus, cross-sectional areas, and moments of inertia. Once the structured problem description is formulated, an additional instruction is appended to indicate the target structural analysis software (e.g., OpenSees, SAP2000, or ETABS). These textual inputs are then provided to the proposed system,which automatically generates executable structural analysis scripts. To account for the inherent stochasticity of LLM generation, each benchmark problem is evaluated through ten independent runs. The reliability is quantified by the proportion of trials that yield error-free analysis scripts.

\section{A Multi-Agent Architecture for Automated Structural Analysis across Software Platforms}
\label{sec:others}

\subsection{Overall architectural design}
\label{sec:overall_system}

\begin{figure*}[htbp]
\centering
\includegraphics[width=0.85\textwidth]{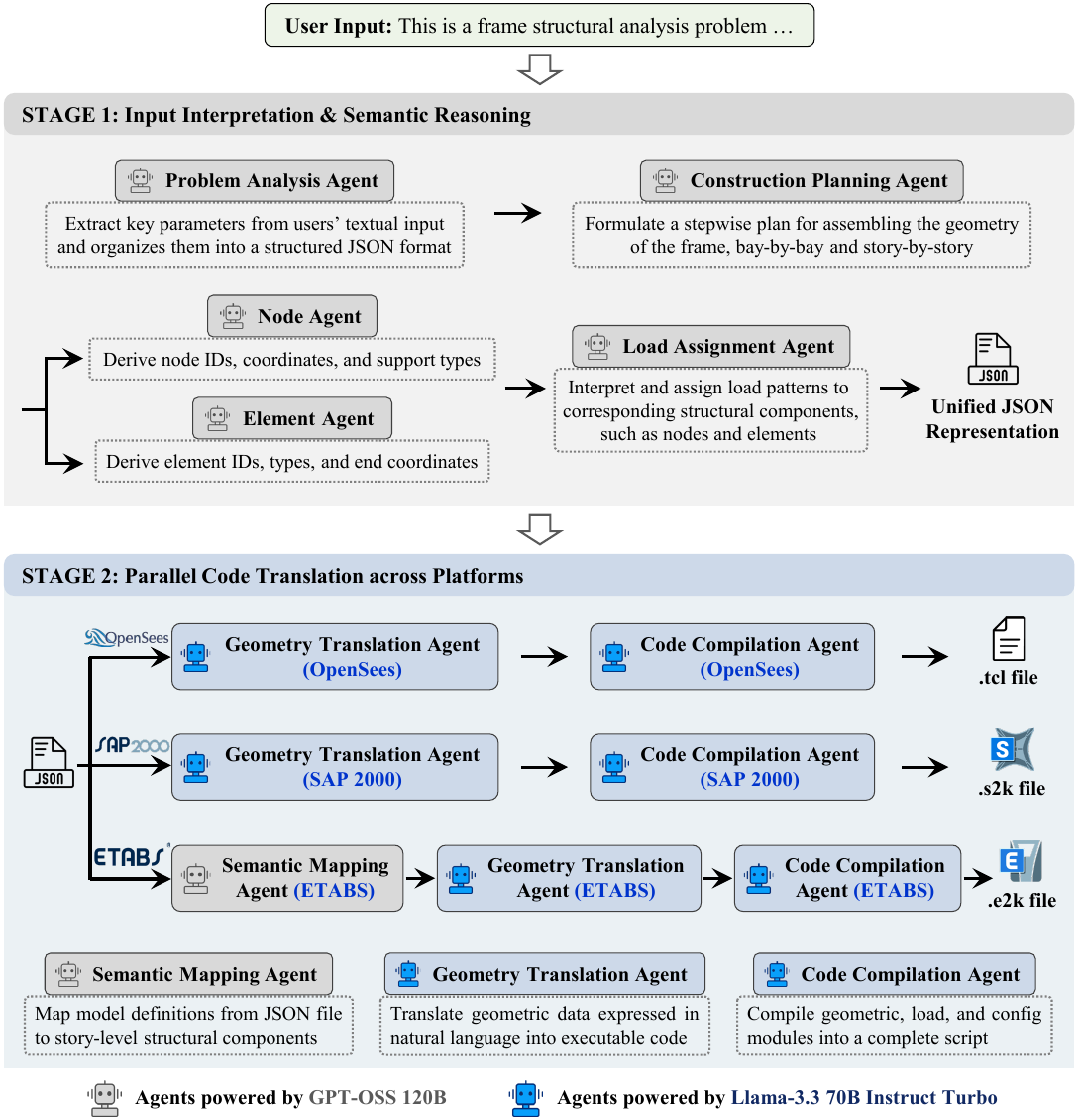}
\caption{Multi-agent system architecture for automated frame structural analysis across FEA platforms.}
\label{Figure4}
\end{figure*}

\cref{Figure4} illustrates the overall architecture design of the proposed multi-agent LLMs to automate structural analysis across multiple software platforms. The architecture has two stages that decouples high-level semantic reasoning from platform-specific code generation. Specifically, Stage 1 focuses on input interpretation and semantic reasoning. The process begins with the problem analysis agent and the construction planning agent, which extract key modeling parameters from the users’ problem description and formulate a stepwise assembly logic for the frames. Subsequently, the node agent and element agent derive node coordinates, support conditions, and element connectivity, whereas the load assignment agent interprets and assigns load patterns to the corresponding structural components. The backbone LLM underpinning these agents is GPT-OSS 120B model, selected for its strong semantic reasoning capability \citep{agarwal2025gpt, bi2025gpt}. Collectively, these agents produce a unified and platform-agnostic JSON representation that encapsulates all information required for finite element modeling of frame structures. 

Stage 2 performs parallel code translation from the unified JSON representation to multiple structural analysis platforms, including OpenSees, SAP2000, and ETABS. For OpenSees and SAP2000, the geometry translation agent converts the structured geometric definitions into executable commands consistent with each platform’s modeling syntax. Then, the code compilation agent assembles geometric, boundary, load, and configuration modules into complete and executable scripts. These agents are powered by the Llama-3.3 70B-Instruct Turbo model due to its robust instruction-following performance \citep{dong2025revisiting, yan2025codeif}. In contrast, ETABS requires an additional simple semantic mapping agent to bridge the gap between generic structural definitions and the modeling logic inherent to the software itself. This agent is powered by GPT-OSS 120B model due to its emphasis on semantic reasoning task. The outputs of this stage are executable code files tailored to each software (e.g., .tcl, .s2k, and .e2k), thereby enabling automated structural analysis across software platforms.

\subsection{Stage 1: Interpretation and reasoning}
\label{sec:stage1}

\begin{figure*}[htbp]
\centering
\includegraphics[width=0.85\textwidth]{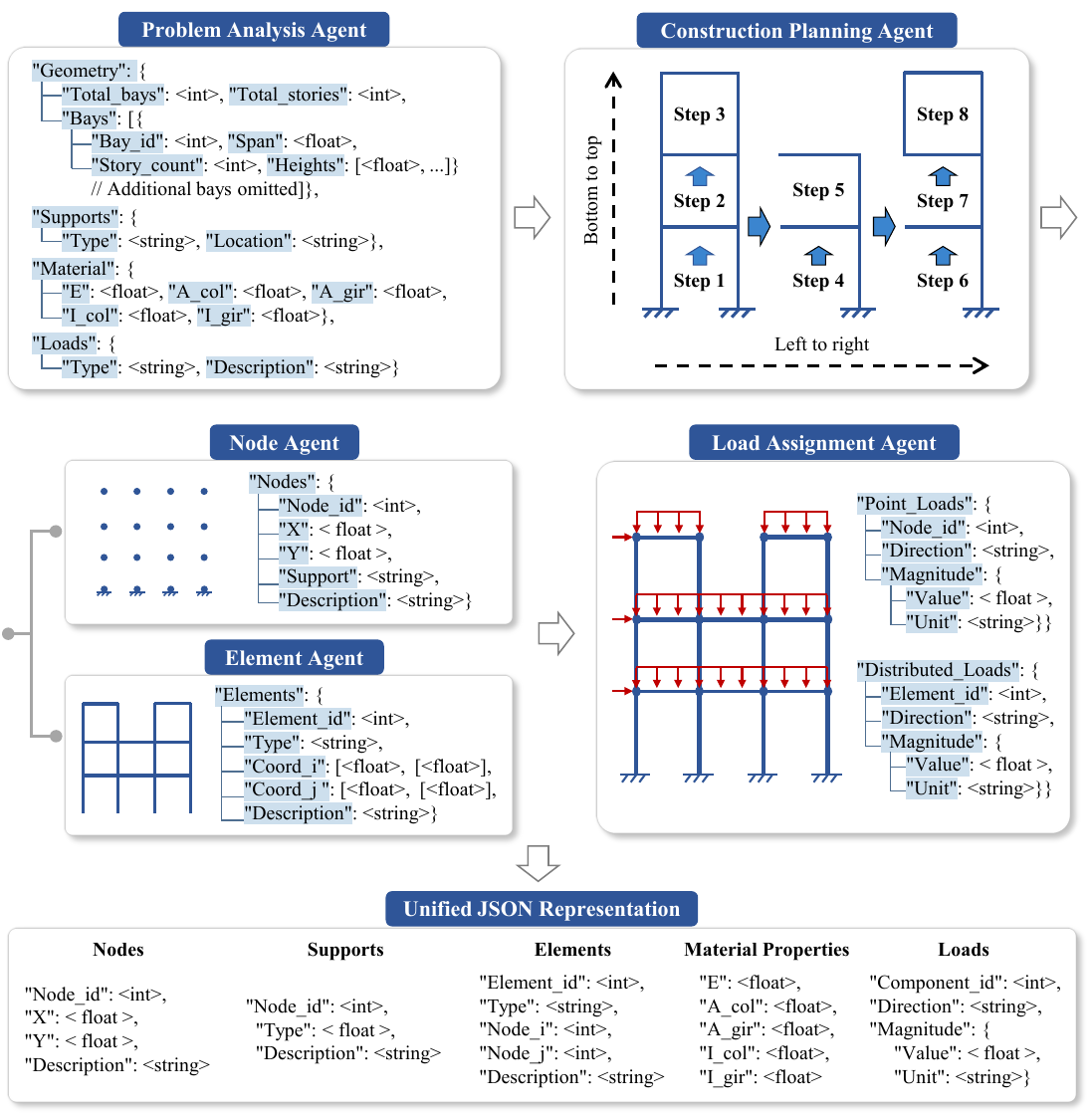}
\caption{Workflow of Stage 1, where multiple agents interpret user input and perform semantic reasoning to generate a unified JSON representation for FEA analysis.}
\label{Figure5}
\end{figure*}

Stage 1 transforms users’ problem description into a structured representation that can be directly used by FEA workflows. As illustrated in \cref{Figure5}, this stage consists of five specialized agents: a problem analysis agent, a construction planning agent, a node agent, an element agent, and a load assignment agent. Specifically, the problem analysis agent performs semantic parsing to extract key structural parameters from natural language input. These parameters include geometry, supports, material properties, and loads. The geometry section records the global structural configuration, such as the total number of bays and stories, as well as bay-level attributes including span lengths, story counts, and story heights. The support section captures the type and location of boundary conditions, including fixed, pinned, and roller supports. The material section stores mechanical and sectional properties, including Young's modulus, cross-sectional area, and moment of inertia. Multiple values are permitted for each parameter, allowing distinct material and sectional properties to be assigned to individual structural members. The load section identifies load types, such as concentrated or distributed, and their application locations. All extracted parameters are organized into a structured JSON format to support downstream processing.

Following parameter extraction, a construction planning agent infers a stepwise assembly sequence using a bottom-to-top, left-to-right heuristic. This sequence provides a deterministic topology assembly logic, ensuring that the structural model is constructed in a consistent and well-defined order. Building upon the inferred assembly logic, the node agent and element agent operate in parallel to incrementally construct the structural geometry. At each construction step, the node agent determines the required nodes and computes their spatial coordinates based on bay spans and story heights, while assigning supports in accordance with the specified boundary conditions. In parallel, the element agent defines structural members, such as columns and girders, and establishes their connectivity by linking the corresponding end nodes. After completing all construction steps, the node agent outputs a JSON file containing node identifiers, spatial coordinates, supports, and descriptions, whereas the element agent produces a JSON file that records element identifiers, types, end node coordinates, and descriptions.

Subsequently, the load assignment agent maps external forces onto the generated structural geometry by assigning concentrated and distributed loads to the appropriate nodes and elements. Each load entry records the target identifier, load direction, magnitude, and unit. Finally, the outputs of all agents are integrated into a unified, software-agnostic JSON representation that encodes nodes, supports, elements, material properties, and loads. To ensure compatibility with standard FEA modeling conventions, an auxiliary Python function is defined to convert element connectivity from end node coordinates to end node identifiers. Together, these steps complete the interpretation and reasoning process required for FEA analysis of frame structures.

\subsection{Stage 2: Parallel code translation}
\label{sec:stage2}

Stage 2 translates the unified JSON representation generated in Stage 1 into executable structural analysis scripts across multiple FEA platforms. As demonstrated in \cref{Figure4}, this stage adopts a multi-agent architecture comprising a geometry translation agents, a code compilation agent, and, when required, a semantic mapping agent. 

\begin{figure*}[htbp]
\centering
\includegraphics[width=0.85\textwidth]{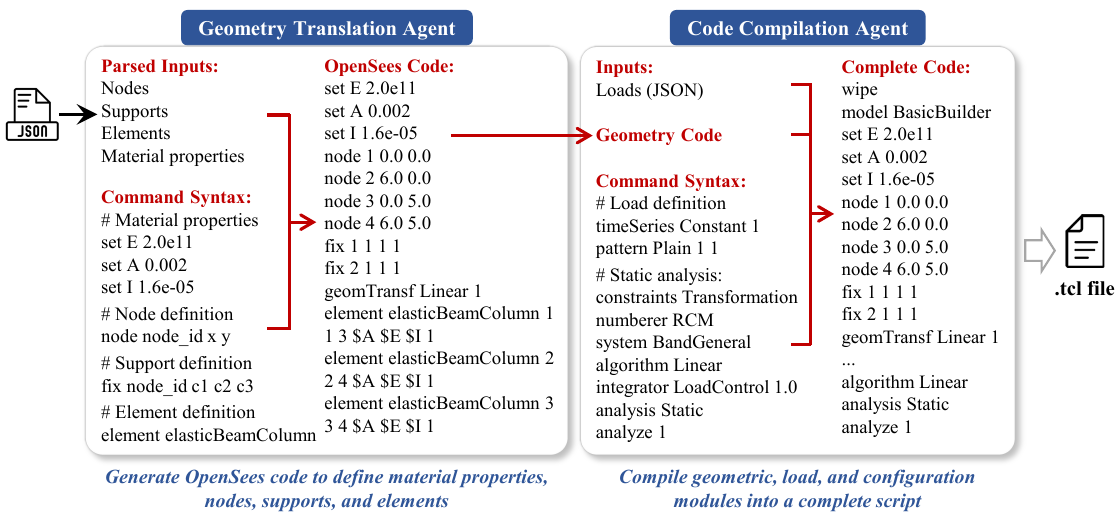}
\caption{Workflow of Stage 2, where geometry translation agent and code compilation agent transform the JSON representation into an executable structural analysis script.}
\label{Figure6}
\end{figure*}

The use of OpenSees and SAP2000 generally involves defining objects such as nodes, elements, material properties, and loads and defines the relations between these objects (e.g., assigning two nodes to a line element). This modeling logic is similar to how the unified JSON representation is set up. Therefore, it is natural to translate each object in the JSON file (e.g., a node, a support, and an element) to a line of code in the script file for OpenSees and SAP2000. Specifically, the translation workflow is performed by two sequential agents, as shown in \cref{Figure6}. First, the geometry translation agent converts the geometric information encoded in the JSON file, including nodes, supports, elements, and material properties, into modeling commands consistent with the target platform. This process produces an intermediate code segment that defines the structural geometry. Subsequently, the code compilation agent translates load definitions from the JSON file into corresponding commands and integrates geometric, load, and analysis configuration modules into a complete executable script. The resulting files can be directly imported and executed in the target platforms, such as .tcl files for OpenSees and .s2k files for SAP2000.

\begin{figure*}[htbp]
\centering
\includegraphics[width=0.85\textwidth]{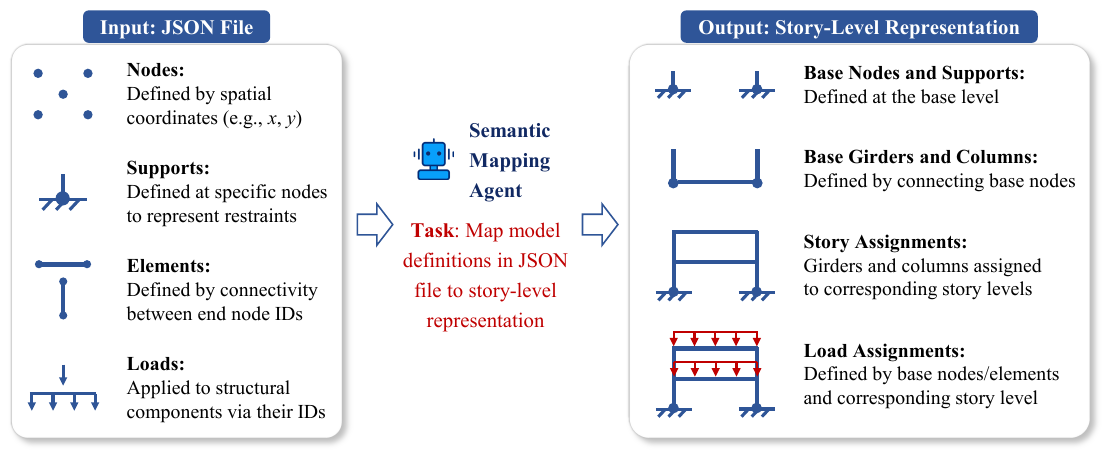}
\caption{Semantic mapping agent that converts JSON files into story-level representations.}
\label{Figure7}
\end{figure*}

ETABS, primarily for building structures, adopts a different modeling logic from OpenSees and SAP2000. It first sets up the stories and then assigns the objects (e.g., columns, beams) to different stories. Specifically, nodes at the base elevation are identified as base nodes, and base elements, including columns and girders, are constructed by establishing connectivity among these nodes. These base elements are assigned to corresponding story levels to form the structural geometry. Similarly, boundary and load conditions are defined by associating them with the appropriate base nodes and elements and assigning them to their respective stories. This modeling logic is different from the that of the JSON representation. Therefore, Stage 2 introduces a semantic mapping agent, as illustrated in \cref{Figure7}, to bridge the gap of modeling logic between the JSON representation and ETABS. Essentially, the semantic mapping agent defines the stories and assign the objects to corresponding stories. Following semantic mapping, the resulting story-level representation is passed to the geometry translation and code compilation agents tailored for ETABS, which generate an executable .e2k file. Notably, Stage 2 adopts a parallel execution strategy, in which translation pipelines for different platforms operate independently. This design not only improves computational efficiency but also facilitates scalability, as it allows additional platforms to be integrated by introducing dedicated translation agents without affecting existing workflows.


\section{Results and Discussion}

\subsection{Performance of the proposed LLMs}

\begin{figure*}[htbp]
\centering
\includegraphics[width=0.85\textwidth]{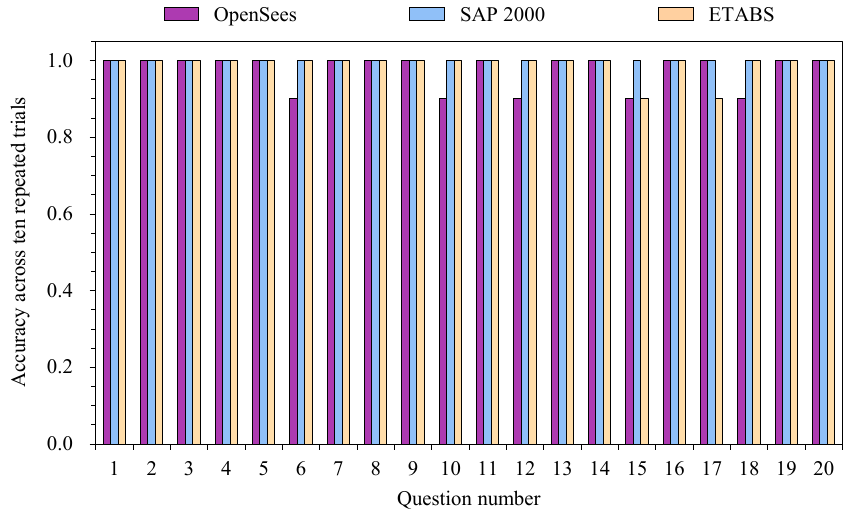}
\caption{Performance of the proposed multi‑agent LLMs across three structural analysis platforms.}
\label{Figure8}
\end{figure*}

The performance of the proposed multi-agent LLMs is evaluated on a set of structural analysis tasks of frames across three widely adopted software platforms, including OpenSees, SAP2000, and ETABS, as illustrated in \cref{Figure8}. For each task, the accuracy is quantified as the success rate over ten repeated trials to account for the stochastic variability inherent in the generation processes of LLMs. Generally, the proposed LLMs achieve consistently strong performance across all platforms, with each question achieving an accuracy exceeding 90\%. These results indicate that the LLMs enable reliable and consistent structural model generation across FEA environments characterized by diverse modeling logic and command syntax. This consistently high accuracy also validates the effectiveness of the underlying design of the multi-agent architecture, in which high-level semantic reasoning is decoupled from script translation for individual FEA platforms.

Among the evaluated platforms, SAP2000 exhibits the highest level of consistency, achieving perfect accuracy for all questions across all trials. Minor performance variations are observed for a small subset of questions, particularly for OpenSees and ETABS. Specifically, ETABS maintains perfect accuracy for 18 out of 20 questions, with slight reductions to an accuracy of 90\% for questions 15 and 17. OpenSees shows relatively greater variability, with accuracy decreasing to 90\% for five questions (6, 10, 12, 15, and 18). These deviations are primarily attributed to differences in syntax requirements across platforms. Importantly, the observed performance variations are limited in magnitude and do not exhibit bias toward any specific platform, underscoring the robustness of the proposed multi-agent architecture. Collectively, these results demonstrate that the proposed LLMs achieve cross-platform consistency and high reliability.

\subsection{Integration with Existing Workflows}

The proposed LLMs are designed to be integrated seamlessly with existing structural analysis workflows by generating executable scripts that can be directly imported into the FEA softwares. Users provide natural language descriptions of structural analysis problems, and the system automatically produces input files compatible with the target platforms. Rather than replacing existing software environments or analysis pipelines, the proposed approach introduces an input interface that complements current engineering practice. This design allows practitioners to retain their expertise with familiar software tools while leveraging automation to reduce manual modeling effort, minimize syntactic errors, and improve overall modeling efficiency.

\begin{figure*}[htbp]
\centering
\includegraphics[width=0.85\textwidth]{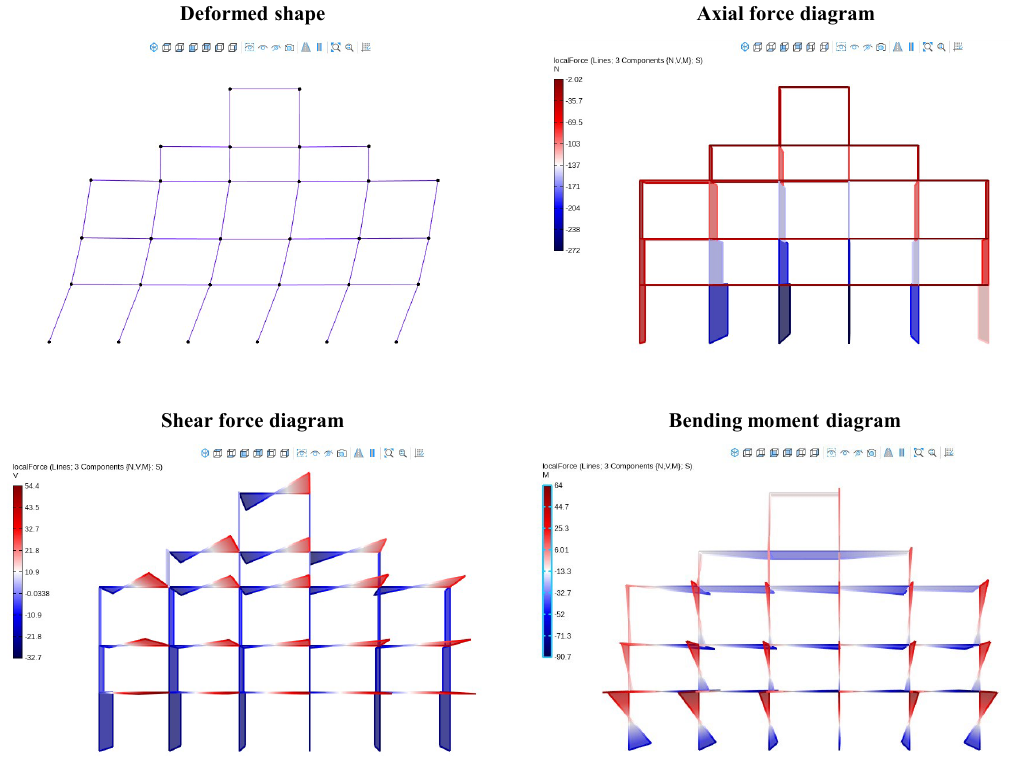}
\caption{Visualization of structural analysis results from OpenSees using STKO.}
\label{Figure9}
\end{figure*}

\begin{figure*}[htbp]
\centering
\includegraphics[width=0.85\textwidth]{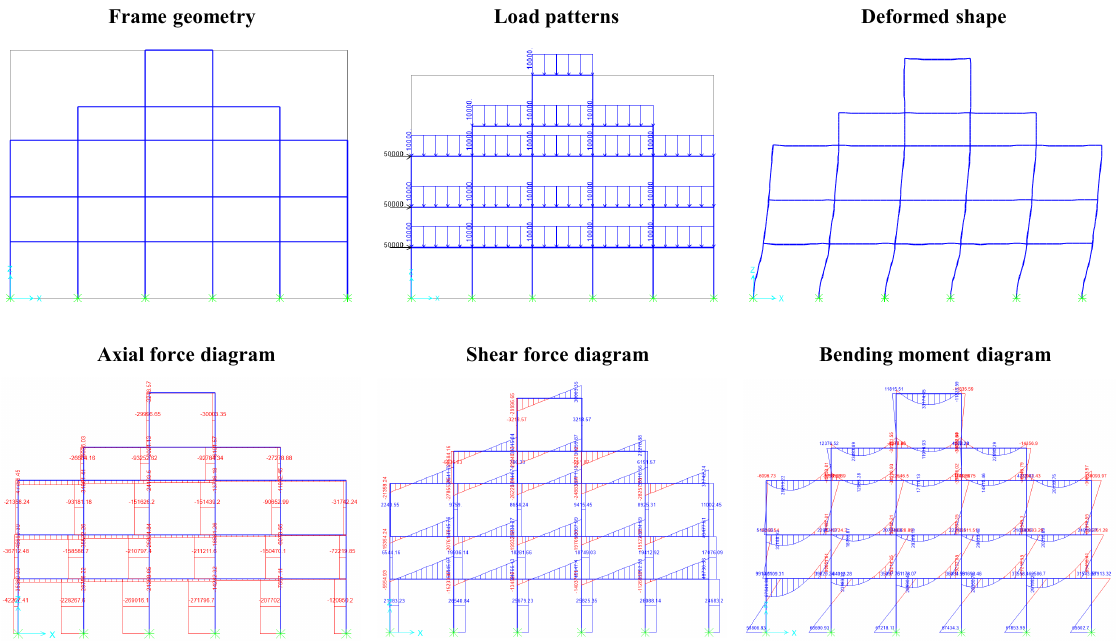}
\caption{Visualization of structural analysis results from SAP2000. }
\label{Figure10}
\end{figure*}

\begin{figure*}[htbp]
\centering
\includegraphics[width=0.85\textwidth]{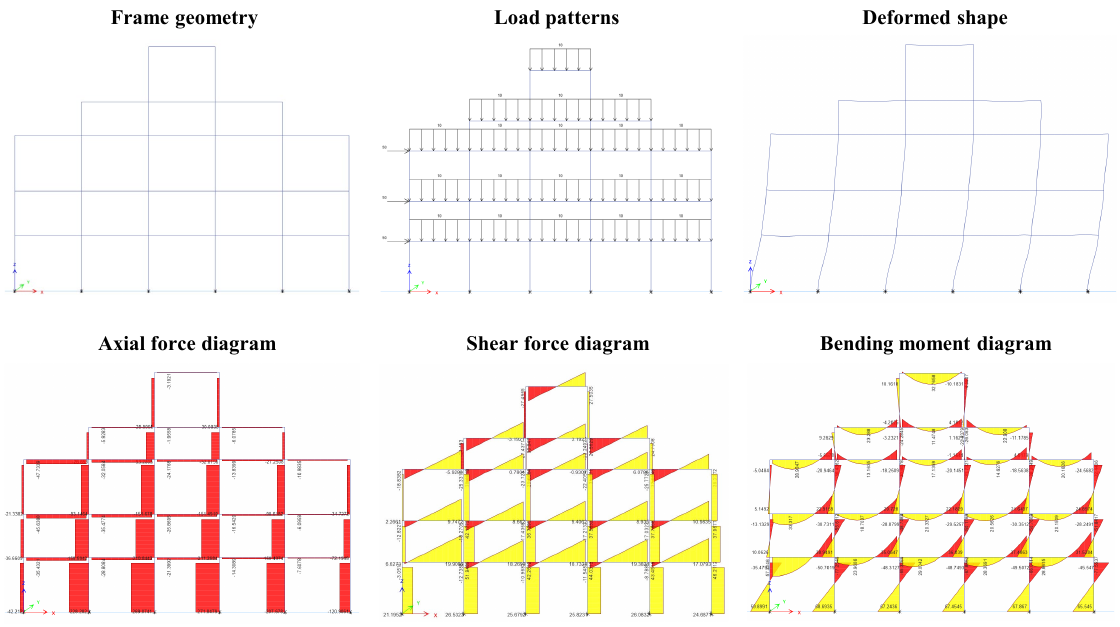}
\caption{Visualization of structural analysis results from ETABS. }
\label{Figure11}
\end{figure*}

For OpenSees, the system generates a .tcl file that can be directly executed for structural analysis. To facilitate result interpretation and post-processing, the analysis outputs are visualized using STKO \citep{stko}, as illustrated in \cref{Figure9}. The visualizations include structural deformation as well as internal force diagrams, such as axial force, shear force, and bending moment diagrams. This integration mitigates the manual workload and steep learning curve associated with TCL scripting. For commercial FEA platforms such as SAP2000 and ETABS, the proposed LLMs output .s2k and .e2k files, respectively, which can be directly imported into the corresponding software. Upon import, the native GUIs allow users to inspect the automatically generated structural models, including geometry, applied loads, boundary conditions, and assigned material properties, as shown in \cref{Figure10} and \cref{Figure11}. After verification, analyses can be executed using the built-in solvers, and results can be visualized in standard forms, including deformed shapes and internal force diagrams. This workflow ensures that engineers remain actively involved in the inspection and validation of structural models, thereby supporting human-in-the-loop decision-making and enhancing transparency in the generated models. Using the proposed architecture, the LLMs further enable direct comparison of structural analysis results across multiple platforms, including OpenSees, SAP2000, and ETABS. Such cross-platform consistency checks are routinely adopted in engineering practice to confirm the robustness of analysis results under different modeling assumptions, thereby enhancing confidence in engineering decisions.

\subsection{Comparison with state-of-the-art LLMs}

To validate the practical value of the proposed LLMs, the performance is benchmarked against two state-of-the-art general-purpose LLMs, namely GPT-5.2 and Gemini-3-Pro. The evaluation involves a two-stage prompting strategy. First, the LLMs are provided with a textual description of a frame structural analysis problem, consistent with the specification format described in \cref{Figure3}. Second, the LLMs are instructed to generate a complete and executable structural analysis script. To minimize ambiguity, the required output format is explicitly specified for each target platform, including .tcl for OpenSees, .s2k for SAP2000, and .e2k for ETABS. For each benchmark problem, ten independent trails are conducted for each of the three target platforms. The generated scripts are imported into the corresponding software environments, where internal force diagrams are produced and compared against the reference solution. An output is considered correct only if the resulting internal force diagrams are consistent with the ground truth. This evaluation protocol provides quantitative insights into the performance of leading general-purpose LLMs within practical structural engineering workflows.

\begin{figure*}[htbp]
\centering
\includegraphics[width=0.85\textwidth]{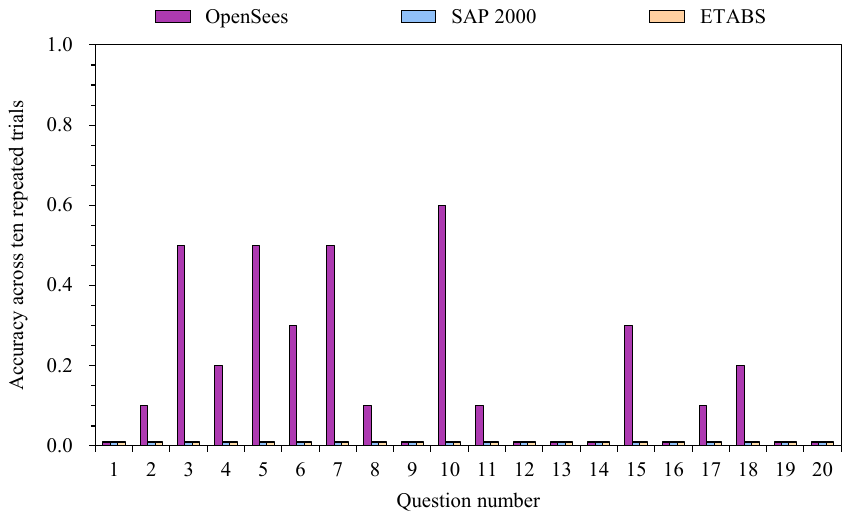}
\caption{Performance of GPT-5.2 on frame structural analysis tasks across three FEA platforms}
\label{Figure12}
\end{figure*}

\begin{figure*}[htbp]
\centering
\includegraphics[width=0.85\textwidth]{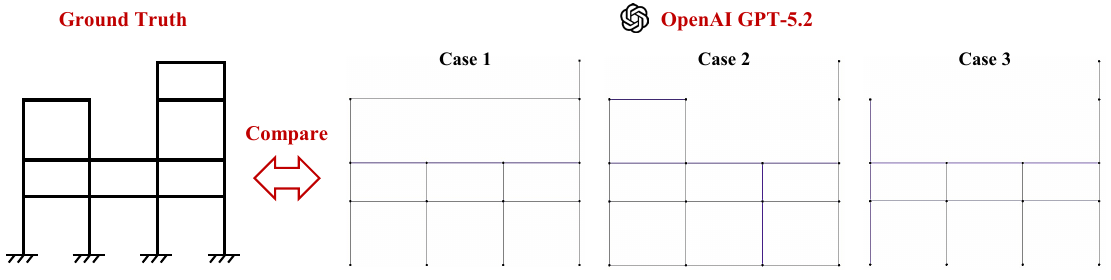}
\caption{Illustrative examples of semantic reasoning errors in OpenSees scripts generated by GPT-5.2.}
\label{Figure13}
\end{figure*}

The performance of GPT-5.2 on the benchmark dataset is illustrated in \cref{Figure12}. Generally, GPT-5.2 demonstrates limited capability in generating correct structural analysis scripts. The average accuracy across 20 problems is 18\% for OpenSees, while performance on SAP2000 and ETABS is 0\%. For OpenSees, three key observations are identified. First, the performance exhibits significant variability across problems. While certain cases such as question 10 achieve accuracy of up to 60\%, most problems yield zero accuracy across all ten trials. This instability highlights a high risk of hallucinations when LLMs are used for code generation in structural analysis workflows. Second, the successful rate of executing the generated OpenSees scripts reaches 51\%. This suggests that GPT-5.2 possesses syntactic knowledge of this open-source platform. Scripts that fail to execute are primarily affected by (i) duplicate node definitions, (ii) undefined or missing elements, and (iii) incompatible variable formats. Third, a notable gap exists between executability (51\%) and correctness (18\%). This discrepancy indicates that, although GPT-5.2 can generate syntactically valid commands, it struggles with the semantic reasoning that translates abstract structural descriptions into detailed modeling definitions. \cref{Figure13} shows these limitations using a 3-2-4 frame as a illustrative example. In case 1, a beam is incorrectly defined to span from the leftmost to the rightmost column, neglecting the reduced story count in the second bay. Case 2 shows an error in which elements are incorrectly connected between support nodes. Case 3 omits critical nodes required to assemble the structure, resulting in an incomplete structural model.


For the commercial software platforms SAP2000 and ETABS, GPT-5.2 fails on all benchmark problems. These failures attribute to the rigorous syntax requirements and specific modeling logics that the general-purpose LLM cannot consistently capture. In the case of SAP2000, errors arise from two primary sources. First, GPT-5.2 frequently produces non-canonical table names. For example, the correct table identifier TABLE: "LOAD PATTERN DEFINITIONS" is often replaced by TABLE: "LOAD PATTERNS". Because SAP2000 relies on strict string matching during file import, such minor deviations prevent key definitions from being recognized and result in immediate import errors. Second, SAP2000 scripts require a two-step procedure for element definition, in which element connectivity is defined first and section properties are assigned subsequently. GPT-5.2 commonly merges these steps into a single definition, producing syntactically invalid or unreadable scripts. For ETABS, the failure mode reflects a fundamental incompatibility in modeling logic. Specifically, ETABS adopts a story-based modeling logic, where base objects (nodes and elements) are first defined and then assigned across stories through explicit story definitions. In contrast, GPT-5.2 persists in using a logic similar to that of OpenSees, explicitly defining each node and element in global coordinates. This mismatch in modeling logic prevents the successful import of generated .e2k files.

\begin{figure*}[htbp]
\centering
\includegraphics[width=0.85\textwidth]{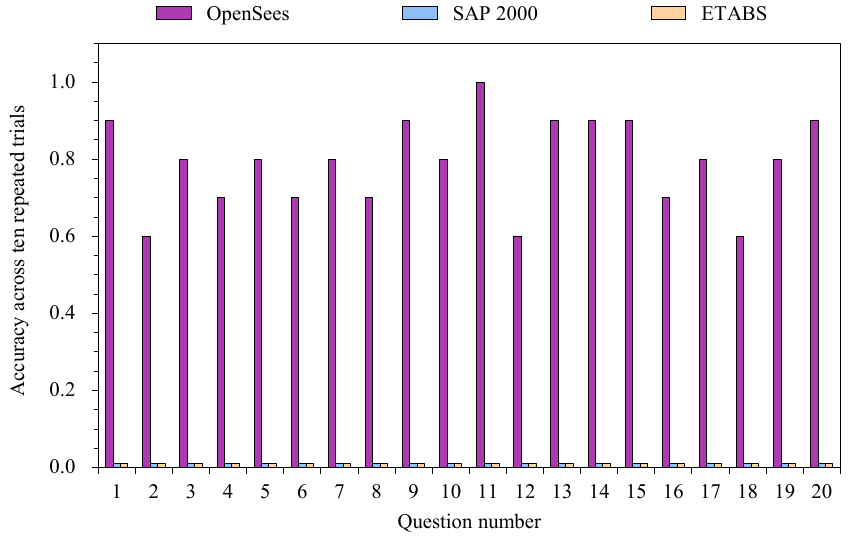}
\caption{Performance of Gemini-3-Pro on frame structural analysis tasks across three FEA platforms}
\label{Figure14}
\end{figure*}

The performance of Gemini-3-Pro is better than GPT-5.2 in using OpenSees, as shown in \cref{Figure14}. Gemini-3-Pro achieves an average accuracy of 79\% across the 20 benchmark problems. Notably, results also show that all executable scripts produce correct internal force diagrams. This consistency suggests that Gemini-3-Pro possesses strong semantic reasoning capabilities in translating structural descriptions into definitions of geometry, materials, loads, and boundary conditions. For the small subset of OpenSees scripts that fail to execute, errors are primarily from syntactic issues, including (i) invalid command names, (ii) duplicate node or element definitions, and (iii) incompatible variable formats. In contrast, Gemini-3-Pro fails on all benchmark problems for the commercial platforms SAP2000 and ETABS. Similar to GPT-5.2, these failures stem from the strict syntax requirements and modeling logic. For SAP2000, although table headers are defined correctly, frequent syntax errors occur in the definitions of material properties, node coordinates, and distributed loads. In particular, it hallucinates for variable names. For example, it replaces XorR, Y, and Z with GlobalX, GlobalY, and GlobalZ, which violates SAP2000’s input grammar and leads to import failures. For ETABS, Gemini-3-Pro exhibits the same fundamental logical incompatibility observed in GPT-5.2. Together, these results indicate that leading general-purpose LLMs remain unreliable for end-to-end structural modeling across FEA platforms. This limitation underscores the practical value of the LLMs developed in this study.

\subsection{Runtime and costs}

To further evaluate the practical feasibility of the proposed multi-agent LLMs, the computational runtime and associated costs are examined. These statistics are collected over the 20 benchmark problems across ten repeated trials. Specifically, the average runtime required to generate structural analysis scripts is 113.54 seconds for OpenSees, 127.56 seconds for SAP2000, and 144.41 seconds for ETABS, respectively. The relatively shorter runtime for OpenSees can be attributed to its concise command syntax, whereas the longer runtime observed for ETABS primarily results from the additional semantic mapping agent. Compared with general-purpose LLMs, the LLMs in this study exhibit a higher runtime. For reference, GPT-5.2 generates scripts with an average runtime ranging from 32.80 to 43.94 seconds, while Gemini-3-Pro requires 80.61 to 94.95 seconds. However, as demonstrated in Section 4.3, these shorter runtimes do not guarantee the reliability or correctness of the generated scripts. From the perspective of practical engineering, the absolute runtime of the proposed LLMs remains highly efficient. The executable scripts for OpenSees, SAP2000, and ETABS can be generated in parallel within approximately two minutes. This capability substantially reduces manual modeling effort and provides engineers with flexibility for multi-platform structural analysis.

In addition to runtime efficiency, the economic feasibility of the proposed LLMs is evaluated based on token consumption and API pricing. While the multi-agent architecture incurs higher token usage due to inter-agent communication, the LLMs are cost-effective by leveraging lightweight open-source backbone models including GPT-OSS-120B and Llama-3.3-70B-Instruct-Turbo. According to the API pricing provided by Together AI (in 2026 US\$), GPT-OSS-120B is priced at \$0.15 per million input tokens and \$0.60 per million output tokens, while Llama-3.3-70B-Instruct-Turbo is priced at \$0.88 per million tokens for both input and output. Under these rates, the total cost of the proposed LLMs range from \$0.012 to \$0.025 per benchmark problem per platform. For comparison, the costs of state-of-the-art general-purpose LLMs are also calculated. GPT-5.2 is priced at \$1.75 per million input tokens and \$14.00 per million output tokens, while Gemini-3-Pro is priced at \$2.00 per million input tokens and \$12.00 per million output tokens. Under the same benchmark settings, these models consume approximately \$0.05 and \$0.04 per problem, respectively. These results demonstrate that the proposed multi-agent LLMs deliver a cost-effective solution for frame structural analysis, exhibiting strong potential for scalability in real-world engineering applications.

\section{Summary and Conclusions}

This paper proposes a multi-agent architecture for large language models (LLMs) to automate frame structural analysis across software platforms. The LLMs adopt a two-stage architecture. In the first stage, five specialized agents, including problem analysis agent, construction planning agent, node agent, element agent, and load assignment agent, are employed to reason over the information required for structural modeling from user inputs. Specifically, the problem analysis agent extracts key parameters from user input and organizes them into a structured JSON format. The construction planning agent formulates a stepwise plan for assembling the frame geometry. The node and element agents derive node coordinates and element connectivity to construct the structural topology. The load assignment agent interprets and assigns load patterns to the corresponding structural components. All information generated in Stage 1 is compiled into a unified JSON representation. In the second stage, the JSON file is translated in parallel into executable scripts for multiple platforms, including OpenSees, SAP2000, and ETABS. For OpenSees and SAP2000, this process is accomplished through a geometry translation agent and a code compilation agent. The geometry translation agent converts geometric and topological information into code commands, while the code compilation agent integrates geometric, loading, and configuration modules into a complete executable script. For ETABS, an additional semantic mapping agent is introduced to transform object-based model definitions into the story-based representation required by the software. The proposed multi-agent architecture leverages lightweight LLM backbones tailored to different stages of the workflow, employing GPT-OSS-120B for semantic reasoning in Stage 1, and Llama-3.3-70B-Instruct-Turbo for code generation in Stage 2. The performance of the proposed LLMs is evaluated using a benchmark dataset comprising 20 representative frame structural analysis problems. The main findings are summarized below:

\begin{itemize}
    \item The proposed multi-agent LLMs demonstrate highly reliable and consistent performance across the benchmark dataset. Over ten repeated trials, each benchmark problem achieves an accuracy exceeding 90\%. The average accuracies across the 20 benchmark problems reach 98\% for OpenSees, 100\% for SAP2000, and 99\% for ETABS, confirming the robustness of the LLMs across software platforms.

    \item The proposed LLMs substantially outperform leading general-purpose LLMs, including GPT-5.2 and Gemini-3-Pro. For open-source platform OpenSees, the average accuracies of GPT-5.2 and Gemini-3-Pro are 18\% and 79\%, respectively, while the proposed LLMs are 98\%. For commercial platforms SAP2000 and ETABS, both general-purpose models fail across all benchmark problems due to their inability to follow rigorous syntax requirements and specific modeling logics, while the proposed LLMs are more than 99\%.
    
    \item The proposed LLMs deliver an efficient and economically viable solution for automated structural analysis. Executable scripts can be generated in parallel within approximately two minutes, while the cost remains low, ranging from \$0.012 to \$0.025 per benchmark problem per platform. This balance of accuracy, efficiency, and cost highlights the practical value of the proposed LLMs for real-world structural engineering workflows. 
    
    \item Despite the strong performance of the proposed LLMs, the current implementation is limited to rectangular frame configurations with vertical columns and horizontal beams. Future work should extend the scope to support complex structural typologies such as bracing systems and cantilevered members. Additionally, the present evaluation relies on a predefined textual description template. Future work should incorporate more real-world use cases to assess the system’s robustness, generalizability, and functional coverage.
    
\end{itemize}

\section*{Data Availability Statement}
Some or all data, models, or code that support the findings of this study are available from the corresponding author upon reasonable request.

\bibliographystyle{ascelike}  
\bibliography{references}




\end{document}